\documentclass[aps,pre,groupedaddress,twocolumn,showpacs]{revtex4}

\usepackage{graphics}
\usepackage{amsmath}
\usepackage{amsfonts}
\usepackage{amssymb}
\usepackage{fancyhdr}
\usepackage{psfrag}

\renewcommand{\vec}[1]{\mbox{\boldmath$#1$}}
\newcommand{\pr}{\|}

\newcommand{\uc}{\xi}

\newcommand{\vek}[1]{#1}

\newcommand{\laplace}{{\vec{\nabla}_\pr^2}}

\begin{document}

\title{Mechanics of non-planar membranes with force-dipole activity}

\author{Michael A. Lomholt}
\email[]{mlomholt@memphys.sdu.dk}
\affiliation{NORDITA - Nordic Institute for Theoretical Physics, Blegdamsvej 17, 2100 Copenhagen \O, Denmark}
\affiliation{The MEMPHYS Center for Biomembrane Physics, University of Southern Denmark, Campusvej 55, DK-5230 Odense M, Denmark}

\date{\today}

\begin{abstract}
A study is made of how active membrane proteins can modify the long wavelength mechanics of fluid membranes. The activity of the proteins is modelled as disturbing the protein surroundings through non-local force distributions of which a force-dipole distribution is the simplest example. An analytic expression describing how the activity modifies the force-balance equation for the membrane surface is obtained in the form of a moment expansion of the force distribution. This expression allows for further studies of the consequences of the activity for non-planar membranes. In particular the active contributions to mechanical properties such as tension and bending moments become apparent. It is also explained how the activity can induce a hydrodynamic attraction between the active proteins in the membrane.
\end{abstract}


\maketitle


\section{Introduction}
Biological membranes actively participate in many biological processes. They constantly exchange material with their surroundings and they contain molecular machines that consume free energy to perform different tasks \cite{alberts02}. Biological membranes are therefore not just simple self-assembled structures of lipids and proteins in thermal equilibrium, but should be considered as non-equilibrium systems.

Micropipette experiments have been performed on lipid model membranes with membrane proteins that actively pump ions across the membrane \cite{manneville99,girard05}. They have demonstrated that the activity of the pumps influence the mechanical properties of the membrane. The membrane protein investigated in \cite{manneville99}, bacteriorhodopsin, was a light-driven proton pump, while the protein studied in \cite{girard05} was the ATP-driven calcium pump ${\rm Ca}^{2+}$-ATPase. In both micropipette experiments it was found that the membrane tension was less sensitive to changes in the visible area of the membrane when the pumps were active. This was interpreted as the shape fluctuations of the membranes being enhanced by the activity. Therefore the results were presented as increased effective temperatures, which could be about two to three times the values found when the micropipette experiments were performed without the pumps being activated.

In \cite{manneville01} a theoretical model was proposed which explains the enhancement of the fluctuations as being a result of the active proteins pushing on their surroundings. The pushing was modelled in the simplest possible way, as a force-dipole acting on the fluids surrounding the membrane. A force monopole was excluded because it would require an external source of force in the system. Completing the model by setting up stochastic equations of motion for the membrane shape to first order in deviations from a planar membrane, including an averaged contribution of the active force-dipoles together with thermal noise in the description, it was shown that this model provides a possible explanation of the enhancement of the fluctuations. However, as noted in \cite{gov04}, the non-thermal noise from the activity may also contribute to the enhancement of the fluctuations.

To test the force-dipole model more stringently it would be necessary to perform other experiments on these active membrane systems. One possibility would be to measure the fluctuation spectrum directly using video microscopy as proposed in \cite{pecreaux04}. Another possibility would be to study how the activity modifies the overall average shape of the membrane. For lipid membranes in equilibrium this shape is well studied both theoretically and experimentally, see for instance \cite{seifert97} for a review. A way to make the average shape sensitive to switching on activity would be to prepare the membrane in a state where a small change in control parameters, such as temperature or osmolarity, will result in a sharp transition in shape, for instance from a prolate to an oblate \footnote{As suggested by a referee of this article, the experiment might be made even more sensitive to the protein activity if the dynamics of the bistability of the shape is studied at the point of the prolate-oblate transition in the way that it is studied in \cite{dobereiner96} for the equilibrium situation.}.

To find the predictions of the force-dipole model for such experiments it would be beneficial to be able to work analytically with the model for membrane conformations that are not close to being planar. However, as the force-dipole model was formulated in \cite{manneville01}, this is not easy to do directly, because the equations of motion for the surrounding bulk fluids would then have to be solved in these non-planar geometries including the presence of the force-dipoles.

The purpose of the present paper is to show how this difficulty can be avoided by using a more indirect approach where the formalism developed in \cite{lomholt05} is applied. The idea behind this approach is to develop an idealized mathematical formulation of the force-dipole model where all effects related to the presence of the membrane are assigned to an infinitely thin surface. The space around this surface is filled with the bulk fluids and excess quantities are then attributed to the surface such that the amounts of extensive quantities like energy, momentum, number of molecules etc. are conserved. This idealized formulation will also be called the Gibbs formulation of the model, since the philosophy behind it, applied to a non-equilibrium situation here, is the one that Gibbs developed to treat the thermodynamics of interfaces \cite{Gibbs61}. For the force-dipole model the most important adjustment needed to arrive at its Gibbs formulation is to replace the non-local contribution of the force-dipoles in the bulk fluid momentum conservation laws with an excess current of momentum on the membrane surface.

The outline of the presentation of this approach is as follows. We will give a brief review of differential geometry of surfaces in section \ref{sec:geom} to establish the notation. Then we will continue with formulating a generalized version of the force-dipole model of \cite{manneville01} in section \ref{sec:model}. This formulation of the model will be called the semi-microscopic formulation. Microscopic, because molecular length scales enter directly, namely the lengths over which the forces are distributed. Semi is added in front because, apart from these activity induced forces, the membrane is treated as being infinitely thin. In section \ref{sec:excess} we then show how an expansion can be performed in these molecular length scales divided by the curvature radii of the membrane to arrive at the Gibbs formulation for the model. In the Gibbs formulation some of the physical consequences of the force-dipole model become clearer than in the semi-microscopic formulation. One of these consequences is an influence of the activity on the tension of the membrane, which is discussed in section \ref{sec:tension}. Another consequence of the force-dipole model, discussed in section \ref{sec:diffusion}, is a hydrodynamic interaction force between the membrane proteins induced by the activity, offering a possible explanation of two experiments reported in \cite{kahya02} on the clustering and diffusion of active bacteriorhodopsin molecules. A conclusion is given in section \ref{sec:conclusion}, and finally an appendix is added where the bulk hydrodynamics is solved directly for the semi-microscopic formulation in the case of a nearly planar membrane. Comparing this solution with the Gibbs formulation derived in section \ref{sec:excess} it can be seen directly that the Gibbs formulation and semi-microscopic formulation are equivalent in this case.

The Gibbs formulation derived here will be applied in another paper \cite{lomholt05d}, where the fluctuation spectrum of a quasi-spherical vesicle with active proteins acting as force-dipoles is derived.

\section{Brief review of differential geometry of surfaces}
\label{sec:geom}
In this section we briefly review the mathematical language of two-dimensional differential geometry, which will be used throughout the rest of the paper.

The dynamic shape of the surface is represented by a space-vector function $\vec{R}=\vec{R}(\uc^1,\uc^2,t)$, where the variables $\uc^1$ and $\uc^2$ parametrize the surface and $t$ represents time. At each point on the membrane surface we have a basis for three-dimensional vectors consisting of two tangential vectors
\begin{equation}
\vec{t}_\alpha \equiv \partial_\alpha\vec{R} \equiv \frac{\partial \vec{R}}{\partial \uc^\alpha}\ ,
\end{equation}
where $\alpha=1,2$, and a unit vector normal to the surface,
\begin{equation}
\label{eq:normaldef}
\vec{n} \equiv \frac{\vec{t}_1\times\vec{t}_2}{|\vec{t}_1\times\vec{t}_2|}\ .
\end{equation}

The metric tensor of the surface is defined by
\begin{equation}
g_{\alpha\beta} \equiv \vec{t}_\alpha\cdot\vec{t}_\beta\;.
\end{equation}
It has an inverse, $g^{\alpha\beta}$, which satisfies
\begin{equation}
g^{\alpha\beta}g_{\beta\gamma} = \delta^\alpha_\gamma\;,
\end{equation}
where $\delta^\alpha_\gamma$ is the Kronecker delta and where the repeated Greek superscript-subscript
indices imply summation following the Einstein summation convention. The metric tensor and its inverse
are used to raise and lower Greek indices as in the following example:
\begin{equation}
\vec{t}^\alpha=g^{\alpha\beta}\vec{t}_\beta\;,\quad \vec{t}_\alpha=g_{\alpha\beta}\vec{t}^\beta\;.
\end{equation}

The curvature tensor $K_{\alpha\beta}$ is
\begin{equation}
K_{\alpha\beta} = \vec{n} \cdot \partial_\alpha\partial_\beta\vec{R} \;.
\end{equation}
From $K_{\alpha\beta}$ the scalar mean curvature $H$ and Gaussian curvature $K$ can be obtained:
\begin{align}
H&=\frac{1}{2}g^{\alpha\beta}K_{\alpha\beta}\ ,\\
K&=\det g^{\alpha\beta}K_{\beta\gamma}\ .
\end{align}

The expression for covariant differentiation of for instance a vector $\vec{w}=w^\alpha\vec{t}_\alpha$ is given by
\begin{equation}
D_\alpha w^\beta = \partial_\alpha w^\beta + w^\gamma \Gamma^\beta_{\gamma\alpha}\;,
\end{equation}
where the Christoffel symbols are defined as
\begin{equation}
\Gamma^\gamma_{\alpha\beta} = \frac{1}{2} g^{\gamma\delta} \left(
                           \partial_\beta g_{\delta \alpha} + \partial_\alpha g_{\beta \delta}-\partial_\delta g_{\alpha \beta} \right) \;.
\end{equation}

Finally, the area of a local differential element of the surface is given by
\begin{equation}
dA=\sqrt{g}d\uc^1d\uc^2\;,
\end{equation}
where $g=\det g_{\alpha\beta}$ is the determinant of the metric tensor.

\section{Semi-microscopic formulation}\label{sec:model}
The force-dipole model proposed in \cite{manneville01} states that the important contribution of the activity of the active membrane proteins to the mechanics and dynamics of the membrane shape is that the proteins push on the surrounding bulk fluids. Mathematically, this was formulated by the addition of a source of force $\vec{F}_{\rm act}$ in the equations of motion of the bulk fluids. Due to the microscopic size of lipid-protein membranes we will assume that we are at low Reynolds number where these equations become
\begin{equation}\label{eq:actNS}
\eta \vec{\nabla}^{2} \vec{v}_\pm-\vec{\nabla} p^\pm +\vec{F}^\pm_{\rm act}=0 \;,
\end{equation}
supplemented by the incompressibility condition
\begin{equation}\label{eq:incomp}
\vec{\nabla}\cdot\vec{v}_\pm=0\;.
\end{equation}
Here $\vec{v}_\pm=\vec{v}_\pm(\vec{r},t)$ are the velocities of the bulk fluid at position $\vec{r}$ and time $t$ on the side that $\vec{n}$ points into ($+$) and the other side ($-$) of the membrane, $p_\pm=p_\pm(\vec{r},t)$ are the corresponding pressures, $\vec{F}^\pm_{\rm act}=\vec{F}^\pm_{\rm act}(\vec{r},t)$ are the appropriate restrictions of the active force density $\vec{F}_{\rm act}$ to the $\pm$-side of the membrane and $\eta$ is the viscosity of the fluids.

The specific expression for $\vec{F}_{\rm act}$ will be generalized slightly here compared with \cite{manneville01}. We will write it as
\begin{equation}\label{eq:Fact}
\vec{F}_{\rm act}(\vec{r})=\int_{\rm M} d A\int d h\;F_{\rm act}\vec{n}\delta^3\left(\vec{r}-\left(\vec{R}+h\vec{n}\right)\right)\;.
\end{equation}
The first integral in Equation (\ref{eq:Fact}) is over the area of the membrane ${\rm M}$, and the second integral is along the normal direction to the membrane with $h$ being the distance to the membrane surface. Together with the delta function these integrals can be thought of as a conversion of coordinates from the Cartesian position vector $\vec{r}$ to the membrane related coordinates $(\uc^1,\uc^2,h)$. $F_{\rm act}$ will be taken to be an arbitrary function of $h$ and fields on the membrane, say $F_{\rm act}=F_{\rm act}(h,H,n_{p^+},n_{p^-})$, and it therefore depends implicitly on $(\uc^1,\uc^2)$ and $t$. $n_{p^\pm}$ represent area densities of the active proteins in the membrane, with $+$ and $-$ indicating the two possible orientations of an asymmetric transmembrane protein which is incorporated in a membrane.

\begin{figure}[t]
\centerline{
\resizebox{8cm}{!}
{
  \includegraphics{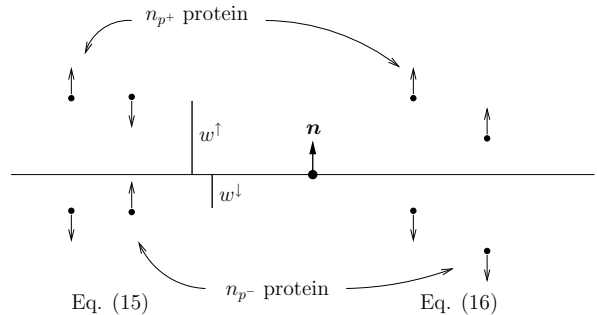}
}}
\caption{Illustration of the points of action and directions of the forces in the force distributions of Equations (\ref{eq:MBRP}) and (\ref{eq:symmMBRP}).}\label{fig:scenarios}
\end{figure}

We can at any point specialize to the more specific force-dipole model of \cite{manneville01} by taking
\begin{equation}
F_{\rm act}=\left(F_a n_\Delta+2H F_a' n_\Sigma\right)\left[\delta\left(h-w^\uparrow\right)-\delta\left(h+w^\downarrow\right)\right]\;,\label{eq:MBRP}
\end{equation}
where $F_a$ and $F_a'$ are constants representing the strength of the active forces and their curvature dependence, $w^\uparrow$ and $w^\downarrow$ are constant lengths giving the distances from the membrane where the forces act. For the protein densities we have introduced $n_\Sigma=n_{p^+}+n_{p^-}$ and $n_\Delta=n_{p^+}-n_{p^-}$.

At this point it should be pointed out that there is a problem with Equation (\ref{eq:MBRP}): it is not invariant with respect to exchanging the two sides of the membrane (see Figure \ref{fig:scenarios}). If locally the physical situation is symmetric with respect to this exchange, i.e. $n_\Delta=0$ and the membrane is planar, then the equation should be invariant with respect to exchanging the two sides. As also noted in \cite{sankararaman02}, this problem can be resolved by using instead the symmetric dipole model
\begin{align}
F_{\rm act}=&\left(F_a + 2 H F_a'\right)n_{p^+}\left[\delta\left(h-w^\uparrow\right)-\delta\left(h+w^\downarrow\right)\right]\nonumber\\
&+\left(F_a - 2 H F_a'\right)n_{p^-}\left[\delta\left(h-w^\downarrow\right)-\delta\left(h+w^\uparrow\right)\right]\;.\label{eq:symmMBRP}
\end{align}
In section \ref{sec:tension} we will discuss what the difference is in the consequences of the two specific models, Equations (\ref{eq:MBRP}) and (\ref{eq:symmMBRP}), for the mechanics of the membrane.

In the Navier-Stokes equation (\ref{eq:actNS}) $\vec{F}_{\rm act}$ acts as a local source term for momentum. Globally, however, there can be no production of momentum. Or equivalently, formulated as in Newtons third law, for every force there has to be an equal but opposite counterforce. This condition will be ensured by assuming that
\begin{equation}
\int d h\;F_{\rm act}=0\;.\label{eq:NewtonIII}
\end{equation}

To complete the description of the bulk hydrodynamics we need to supplement Equations (\ref{eq:actNS}) and (\ref{eq:incomp}) with boundary conditions on the velocities $\vec{v}_\pm$ at the membrane position $\vec{R}$ and at the boundaries of the full membrane-bulk fluid system. We will not choose any specific boundary conditions here, but for later use we will introduce the following notation for the boundary conditions at the membrane
\begin{equation}
\left.\vec{v}_\pm\right|_{\vec{r}=\vec{R}}=\vec{U}_\pm\;,\label{eq:genbound}
\end{equation}
where $\vec{U}_\pm=\vec{U}_\pm\left(\uc^1,\uc^2,t\right)$ are unspecified.

Equations (\ref{eq:actNS}) and (\ref{eq:incomp}) can be solved for the case of an almost planar membrane, expanding to first order in deviations of the membrane shape from a plane. This is carried out in the appendix of this paper. It was also carried out in a different manner in \cite{manneville01}, where the result was used to explain data obtained from micro-pipette experiments on active membranes. However, the large scale geometry of the membranes in micropipette experiments are not planar. Treating them as planar is an approximation justified for studying membrane undulations with short wavelengths compared to the typical length scales of the overall membrane geometry. For the case of micropipette experiments it is sufficient to know the behavior of the short wavelength undulations because the long wavelength undulations are suppressed by the tension that is induced when the membrane is aspirated to the pipette. But as explained in the Introduction there are situations where it is useful to know the behavior of the long wavelength undulations or the overall geometry of the membrane shape. The problem with the semi-microscopic formulation is that directly solving the bulk fluid hydrodynamics analytically for this formulation seems to be quite difficult, if not impossible, for a typical membrane geometry like one being close to a sphere. In the next section we will show how this problem can be circumvented by a more indirect approach to the problem.

\section{Gibbs formulation}\label{sec:excess}
In this section we will derive the Gibbs formulation corresponding to the semi-microscopic formulation of the generalized force-dipole model of the last section.

The designation Gibbs formulation will in this context be taken to mean a mathematical model of the membrane system in which the membrane is treated entirely as being a surface with zero thickness. The semi-microscopic formulation of the generalized force-dipole model in the last section is not a Gibbs formulation, because the active membrane proteins are acting with forces in the bulk fluids across a finite distance transverse to the membrane. Thus in the corresponding Gibbs formulation these active forces will have to be absent from the bulk fluid equations of motion, i.e. instead of Equation (\ref{eq:actNS}) we would have 
\begin{equation}
\eta \vec{\nabla}^{2} \vec{v}_\pm-\vec{\nabla} p^\pm=0\;,\label{eq:noactNS}
\end{equation}
and the effect of the active forces will have to be present in the equations of motion for the membrane surface and the boundary conditions for the bulk fluids instead.

One of the equations of motion for the membrane surface is the conservation of excess momentum. At low Reynolds number where inertial and convective terms can be discarded this law becomes a force balance equation, which can be written \footnote{For actual estimates of Reynolds numbers for membrane hydrodynamics the reader is referred to \cite{miao02}}
\begin{equation}\label{eq:fbal}
\vec{f}_{\rm rs}+\vec{f}_{\rm dis}+\vec{T}^++\vec{T}^-+\vec{f}_{\rm act}=0\;.
\end{equation}
Here $\vec{f}_{\rm rs}$ is the elastic restoring force of the membrane derivable from the membrane excess free energy $F$ by functional differentiation \cite{lomholt05}
\begin{equation}
\vec{f}_{\rm rs}=-\frac{1}{\sqrt{g}}\frac{\delta F}{\delta \vec{R}}\;.
\end{equation}
The free energy for a lipid membrane should include terms representing bending resistance and tension
\begin{equation}\label{eq:Fbend}
F=\int_{\rm M} d A\;\left[2\kappa H^2+\sigma_0+\dots\right]\;,
\end{equation}
where $\kappa$ is the bending rigidity, $\sigma_0$ a tension and the dots represent terms involving free energy variations with respect to changes in other fields such as the density fields $n_{p^\pm}$. The force $\vec{f}_{\rm dis}$ is associated with internal dissipation in the membrane. However, this dissipative force is often taken to be negligible in comparison with the dissipation in the bulk fluids because of small thickness of the membrane. Instead dissipation enters through $\vec{T}^\pm$, which are the forces from the bulk fluids on the membrane. Given that the stress tensors of the bulk fluids are
\begin{equation}
\mathsf{T}^\pm=-p^\pm\mathsf{I}+\eta\left[\vec{\nabla}\vec{v}_\pm+\left(\vec{\nabla}\vec{v}_\pm\right)^T\right]\;,
\end{equation}
where $\mathsf{I}$ is the identity tensor, we can find $\vec{T}^\pm$ as
\begin{equation}\label{eq:Tvecdef}
\vec{T}^\pm=\pm \vec{n}\cdot\mathsf{T}^\pm\;.
\end{equation}
The main character in this paper is $\vec{f}_{\rm act}$, which is the force on the membrane induced by the activity of the membrane proteins. Note that the force balance equation for the membrane in the semi-microscopic formulation would be Equation (\ref{eq:fbal}) without $\vec{f}_{\rm act}$, because in this formulation the activity induced force is included through $\vec{T}^\pm$. Thus in the semi-microscopic formulation the forces acting directly on the membrane is the same as those included in earlier formalistic works like for instance \cite{cai95,miao02}.

Figuring out what the force $\vec{f}_{\rm act}$ is in the Gibbs formulation is, however, not trivial. A priori $\vec{f}_{\rm act}$ can be divided into two contributions
\begin{equation}
\vec{f}_{\rm act}=\vec{\sigma}_{\rm act}+D_\alpha\vec{T}_{\rm act}^\alpha\;,
\end{equation}
where $\vec{\sigma}_{\rm act}$ is the excess source of force per area due to the activity and $\vec{T}_{\rm act}^\alpha$ is the excess stress induced by the activity. $\vec{\sigma}_{\rm act}$ can be found by projecting the volume integral of $\vec{F}_{\rm act}$ down to an area integral on the membrane surface giving
\begin{equation}
\vec{\sigma}_{\rm act}=\int d h\;\left(1-2 h H+h^2 K\right)\vec{F}_{\rm act}\;,
\end{equation}
since the volume element can be written $d V=\left(1-2hH+h^2K\right)d A\,d h$. However, as a consequence of Equation (\ref{eq:NewtonIII}), there is no excess source of momentum
\begin{equation}
\vec{\sigma}_{\rm act}=0\;.
\end{equation}
This leaves us with $\vec{T}^\alpha_{\rm act}$, which can be found by requiring that the stress on surfaces intersecting the membrane orthogonally is the same in the semi-microscopic formulation and the Gibbs formulation. This implies \cite{lomholt05b}
\begin{equation}\label{eq:3Dto2D}
\vec{T}^\alpha_{\rm act}=\int dh\;\left[g^{\alpha\beta}-h \left(2 H g^{\alpha\beta}-K^{\alpha\beta}\right)\right]\vec{t}_\beta\cdot\mathsf{T}_{\rm act}\;,
\end{equation}
where $\mathsf{T}_{\rm act}$ is the bulk fluid stress induced by the activity of the membrane proteins. More precisely, $\mathsf{T}_{\rm act}$ is the bulk fluid stress tensor in the semi-microscopic formulation minus the bulk fluid stress tensor in the corresponding Gibbs formulation, with the boundary conditions for Equations (\ref{eq:actNS}) and (\ref{eq:noactNS}) chosen such that the solutions for the bulk fluid motion are identical (for identical membrane shapes and protein density fields) in the two formulations at distances from the membrane greater than those connected with the activity induced forces \cite{lomholt05}.

It may seem that to find $\vec{T}^\alpha_{\rm act}$ as a function of the fields on the membrane we still need to work out the hydrodynamics in the semi-microscopic formulation completely. However, if we do not insist on obtaining the exact formula for $\vec{T}^\alpha_{\rm act}$, we can obtain an answer as a moment expansion of $F_{\rm act}$ in the distance $h$ from the membrane in a simpler way.

The strategy we will follow is to work out the formula for the moment expansion of $\vec{T}^\alpha_{\rm act}$ up to the second moment in a special case, and then use a combination of symmetry arguments and dimensional analysis to argue that the formula obtained in the special case is actually the general formula. We will also use the linearity of the problem to divide the force $\vec{F}_{\rm act}$ into infinitesimal contributions of the form
\begin{equation}
\vec{F}_{{\rm act},h}=d h \int_M d A \;F_{\rm act}\vec{n}\delta^3\left(\vec{r}-\left(\vec{R}+h\vec{n}\right)\right)\;,
\end{equation}
where $d h$ should be understood as an infinitesimal length. Then we can find the corresponding contribution $\vec{T}^\alpha_{{\rm act},h}$ to $\vec{T}^\alpha_{\rm act}$ for each of these, and the full result is obtained by performing an integral over $h$ in the end.

The special case we will study is that of a homogeneous membrane with constant curvature, such that $n_{p^\pm}$, $H$ and $K$ are constant and $D_\gamma K_{\alpha\beta}=0$. Examples of such geometrical surfaces would be planes, cylinders and spheres. The boundary condition for the Navier-Stokes Equation (\ref{eq:actNS}) is chosen to be that the velocity should vanish at the position of the membrane.

\begin{figure}[t]
\centerline{
\resizebox{8cm}{!}
{
  \includegraphics{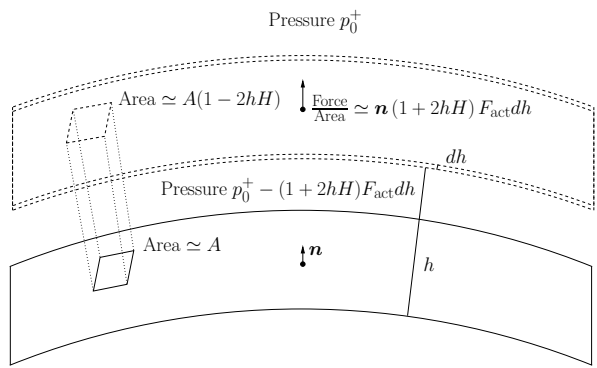}
}}
\caption{Illustration of considerations leading to Equation (\ref{eq:actpress}). Note that $H$ is negative.}\label{fig:thefig}
\end{figure}

The solution to the hydrodynamics in this special case will simply be that the bulk fluid will not move, $\vec{v}_\pm=0$, but there will be a jump in the pressure at a distance $h$ from the membrane to balance the active force. The discontinuous jump for a planar membrane will be $F_{\rm act} d h$, with the lower pressure close to the membrane if $F_{\rm act}\vec{n}$ points away from the membrane. If the membrane is curved with constant curvature, then the force acting at the distance $h$ will be changed slightly, due to the fact that the area of the surface at the distance $h$ differs from the membrane area (see Figure \ref{fig:thefig}). To first order in $h$ we will have a jump in the pressure which is given by
\begin{equation}
\Delta p^\pm_h=\left\{\begin{array}{ll}\mp\left(1+ 2Hh+O(h^2)\right)F_{\rm act} d h\;,\quad&\pm h>0\;,\\
0&\pm h < 0\;.\end{array}\right.\label{eq:actpress}
\end{equation}
This is the change in pressure between the membrane and the distance $h$ induced by the active force $\vec{F}_{{\rm act},h}$. From Equation (\ref{eq:3Dto2D}) we have that the corresponding excess stress, $\vec{T}^\alpha_{{\rm act},h}$ is
\begin{equation}\label{eq:3Dto2Dact}
\vec{T}^\alpha_{{\rm act},h}=\int dh'\;\left(g^{\alpha\beta}-h' \left(2 H g^{\alpha\beta}-K^{\alpha\beta}\right)\right)\vec{t}_\beta\cdot\mathsf{T}_{{\rm act},h}\;,
\end{equation}
where $\mathsf{T}_{{\rm act},h}$ is equal to $-\Delta p^\pm_h$ times the identity for values of $h'$ between $0$ and $h$, and zero otherwise. Performing the integration over $h'$ we get to second order in $h$ for the tangential part $T_{{\rm act},h}^{\alpha\beta}=\vec{T}^\alpha_{{\rm act},h}\cdot\vec{t}^\beta$,
\begin{equation}
T_{{\rm act},h}^{\alpha\beta}=F_{\rm act} h\;d hg^{\alpha\beta}+\left(Hg^{\alpha\beta}+\frac{1}{2}K^{\alpha\beta}\right)F_{\rm act} h^2\;d h\;.
\end{equation}
Finally, performing the integral over $h$ we arrive at
\begin{equation}\label{eq:tangTact}
T_{\rm act}^{\alpha\beta}=\vec{T}^\alpha_{\rm act}\cdot\vec{t}^\beta=\sigma_{\rm dip}g^{\alpha\beta}+\left(Hg^{\alpha\beta}+\frac{1}{2}K^{\alpha\beta}\right)Q\;,
\end{equation}
where
\begin{align}
\sigma_{\rm dip}&=\int d h\; h F_{\rm act}\;,\\
Q&=\int d h\;h^2 F_{\rm act}\;,
\end{align}
are the dipole and quadrupole moment of the force distribution $F_{\rm act}$.

Even though formula (\ref{eq:tangTact}) was derived for homogeneous membranes with constant curvature, it has to be true for any geometry and distribution of active protein molecules, as long as the geometry and distribution are smooth enough such that we can expand $T_{\rm act}^{\alpha\beta}$ in curvature and gradient operators. This is true because other contributions will have to involve the gradient operators. And since we are building a second-rank tensor, there will have to be at least two of them (the only vectors we can use are the gradients of the force distribution and the curvature). From dimensional analysis we can then see that the lowest order quantity we can build using gradient operators involves the gradient used twice on the third moment of $F_{\rm act}$, which is one order higher than what we are expanding to.

The above symmetry argument also tells us that we cannot derive the normal component $T^\alpha_{{\rm n,act}}=\vec{T}^\alpha_{\rm act}\cdot\vec{n}$ directly by using a homogeneous membrane, since there will have to be at least one gradient in the formula for $T^\alpha_{{\rm n,act}}$ to end up with a surface vector. However, there is another way, which goes via obtaining a general formula for the bending moments, $M^{\alpha\beta}_{{\rm act}}$, arising due to the activity. These bending moments are defined such $\vec{R}\times\vec{T}^\alpha_{\rm act}+M^{\alpha\beta}_{\rm act}\vec{n}\times \vec{t}_\beta$ is minus the non-convective part of the excess angular momentum flow induced by the activity, i.e. $M^{\alpha\beta}_{{\rm act}}$ is related to the activity induced internal angular momentum current in the Gibbs formulation. By requiring that the angular momentum flow through surfaces intersecting the membrane orthogonally is the same in the semi-microscopic formulation and the Gibbs formulation one can derive that \cite{lomholt05b}
\begin{equation}\label{eq:microbendact}
M^{\alpha\beta}_{{\rm act}}=\int dh'\ h'\left(g^{\alpha\gamma}-h' \left(2 H g^{\alpha\gamma}-K^{\alpha\gamma}\right)\right)\vec{t}_\gamma\cdot\mathsf{T}_{{\rm act}}\cdot\vec{t}^\beta\;.
\end{equation}
Since there is no excess source of torque in the system, we will have that the internal torque in the membrane vanishes. The excess internal torque is the excess torque on the membrane minus the torque of the excess force on the membrane \cite{lomholt05b}, i.e. its activity induced part is
\begin{equation}
D_\alpha\left(\vec{R}\times\vec{T}^\alpha_{\rm act}+M^{\alpha\beta}_{\rm act}\vec{n}\times \vec{t}_\beta\right)-\vec{R}\times\vec{f}_{\rm act}\;.\label{eq:actindtor}
\end{equation}
The balancing of torques in both the semi-microscopic formulation and the Gibbs formulation implies that the activity induced part of the internal torque vanishes by itself. Equating the tangential part of Equation (\ref{eq:actindtor}) by zero we find that the bending moments are related to the normal component of the stress by the formula
\begin{equation}
{T}^\alpha_{{\rm n,act}}=D_\beta M^{\alpha\beta}_{{\rm act}}\;.
\end{equation}
Inserting the solution for $\mathsf{T}_{{\rm act},h}$ for the homogeneous membrane case in Equation (\ref{eq:microbendact}) and integrating with respect to both $h$ and $h'$, we obtain
\begin{equation}
M^{\alpha\beta}_{\rm act}=\frac{1}{2}Qg^{\alpha\beta}\;.\label{eq:actbend}
\end{equation}
Since the bending moments again form a second rank tensor we can use the same symmetry argument as for $T^{\alpha\beta}_{\rm act}$ to argue that (\ref{eq:actbend}) is the general formula for $M^{\alpha\beta}_{\rm act}$, and we thus have the desired formula
\begin{equation}\label{eq:normTact}
{T}_{\rm n,act}^\alpha=\frac{1}{2}\partial^\alpha Q\;,
\end{equation}
correct to the second moment of $F_{\rm act}$ for any smooth membrane composition and shape.

As briefly touched upon after Equation (\ref{eq:3Dto2D}), we need to have a boundary condition for the bulk fluid in the Gibbs formulation that ensures that we get the same behavior of the bulk fluid far away from the membrane as in the semi-microscopic formulation \cite{lomholt05}.

To find the boundary condition in the Gibbs formulation, we will again study a special case in the semi-microscopic formulation. This time the special case will be a planar membrane positioned at $z=0$ in a Cartesian coordinate system, $(x,y,z)$, but with an inhomogeneous distribution of active proteins, such that the force obeys $F_{\rm act} d h=C x$ with $C$ being independent of $x$ and $y$. For the case of a boundary condition where the velocity vanishes at the membrane we can now solve Equation (\ref{eq:actNS}) when $\vec{F}_{\rm act}$ is restricted to $\vec{F}_{{\rm act},h}$. For the upper part of the bulk fluid a solution is that $p^+$ and $\vec{v}_+$ are constant outside $h$ and for $z<h$ they are
\begin{align}
p^+&={\rm constant}-C x\;,\\
\vec{v}_+&=-\frac{C}{2\eta}z\left(z-2h\right)\vec{\hat{x}}\;.\label{eq:vprofile}
\end{align}
This implies, due to the continuity of $\vec{v}_+$ at $z=h$, that in the rest of the bulk we will have
\begin{equation}
\vec{v}_+=\frac{C}{2\eta}h^2\vec{\hat{x}}\;.\label{eq:specbound}
\end{equation}

The parametrization invariant boundary condition in the Gibbs formulation that agrees with Equations (\ref{eq:specbound}) and (\ref{eq:genbound}) is
\begin{equation}
\left.\vec{v}_\pm\right|_{\vec{r}=\vec{R}}=\vec{U}_\pm\pm\frac{1}{2\eta}\vec{t}^\alpha D_\alpha Q^\pm\;,\label{eq:actbc}
\end{equation}
where
\begin{equation}
Q^\pm=\pm\int_0^{\pm \infty} d h\;h^2F_{\rm act}\;.
\end{equation}
Again we can use symmetry arguments and dimensional analysis to justify that Equation (\ref{eq:actbc}) is the general formula for the boundary condition up to the second moment of $F_{\rm act}$ true for any smooth configuration of the membrane.

Even though the modification of the boundary condition in Equation (\ref{eq:actbc}) is a strict consequence of the generalized force-dipole model formulated in section \ref{sec:model}, one can argue that for some proteins it is an anomaly arising because the model is too simplistic. In the case of for instance bacteriorhodopsin, we have that the physical extension of the proteins outside the width of the lipid membrane is limited \cite{kimura97}. In this case it seems reasonable that an improvement of the Gibbs formulation would be to discard the modification in Equation (\ref{eq:actbc}) and use the original Equation (\ref{eq:genbound}) instead. Whether or not the same argument can be used to question how realistic the modification of the membrane force-balance coming from the force-dipole model, $\vec{f}_{\rm act}$ in Equation (\ref{eq:fbal}), is relies on how close the mechanical properties of the lipids in the membrane resemble those of an incompressible fluid. We will not try to resolve this here.

It was mentioned at the end of section \ref{sec:model} that bulk fluid hydrodynamics in the semi-microscopic formulation can be solved directly for an almost planar membrane. As a check of the correctness of the Gibbs formulation derived in this section, this solution is presented in appendix \ref{sec:planact}. The resulting modification of the force balance equation for the membrane that the active proteins are found to induce are in complete agreement with the results derived in this section.

The momentum conservation law is not necessarily the only conservation law in the Gibbs formulation that is modified through the activity of the proteins. A possible mechanism through which the active proteins can attract each other, leading to modification of their diffusion equations, is discussed in section \ref{sec:diffusion}.

\section{Modification of the tension}\label{sec:tension}
Some of the consequences of the force-dipole model for the mechanics of the membrane emerges readily in the Gibbs formulation. In this section we will discuss the effect of the dipole moment $\sigma_{\rm dip}$ on the tension of the membrane.

From Equations (\ref{eq:tangTact}) and (\ref{eq:normTact}) we see that $\sigma_{\rm dip}$ only contributes to the isotropic part of the tangential stress in the membrane. This part is a two-dimensional analog of pressure in three dimensions \cite{kralchevsky94,capovilla02,lomholt05b}, but due to the difference in sign it is called tension. Thus the role of the dipole moment $\sigma_{\rm dip}$ is exactly that of adding a local contribution to the tension of the membrane.

For the specific dipole model of \cite{manneville01}, Equation (\ref{eq:MBRP}), the active contribution is
\begin{equation}\label{eq:sigmaMBRP}
\sigma_{\rm dip}=\left(w^\uparrow+w^\downarrow\right)\left(F_a n_\Delta+2 H F_a'n_\Sigma\right)\;.
\end{equation}
If $n_\Delta=0$ and the membrane is planar, which they actually go to some length to make certain are reasonable assumptions in the micropipette experiments of \cite{manneville99}, then the prediction of Equation (\ref{eq:sigmaMBRP}) is that $\sigma_{\rm dip}=0$ on average. However, as mentioned in section \ref{sec:model}, the specific model of Equation (\ref{eq:MBRP}) is not realistic because it does not respect the symmetry of exchanging the two sides of the membrane, which should be there when $n_\Delta=0$ and $H=0$. Therefore we will instead use the expression for $\sigma_{\rm dip}$ that one obtains from the symmetric model of Equation (\ref{eq:symmMBRP}), which is
\begin{equation}\label{eq:sigmasymmMBRP}
\sigma_{\rm dip}=\left(w^\uparrow+w^\downarrow\right)\left(F_a n_\Sigma+2 H F_a'n_\Delta\right)\;.
\end{equation}

It should be mentioned that even though the expression for the active contribution to the tension is different for the two dipole models, their predictions for the effective temperature in the micropipette experiments are the same. This is the case because the tension is fixed in the micropipette experiments by other control parameters such as pressures, and because the expression for the other parameter contributing to the mechanics, the quadrupole moment:
\begin{equation}
Q=\left(\left(w^\uparrow\right)^2-\left(w^\downarrow\right)^2\right)\left(F_a n_\Delta+2 H F_a'n_\Sigma\right)\;,
\end{equation}
is identical for the two models.

To give an estimate of the active contribution to the tension, Equation (\ref{eq:sigmasymmMBRP}), we will use an estimate of a quantity $\mathcal{P}_a$ made in \cite{manneville01}, which was found to agree well with the micropipette experiments on bacteriorhodopsin. The quantity is defined as
\begin{equation}
\mathcal{P}_a\equiv F_a\frac{\left(w^\uparrow\right)^2-\left(w^\downarrow\right)^2}{2 w}\;,
\end{equation}
where $w\simeq 5\,{\rm nm}$ is the membrane thickness, and the estimate is
\begin{equation}\label{eq:Paest}
\mathcal{P}_a \simeq 10 k_{\rm B}T\;.
\end{equation}
Making the assumption $w^\uparrow-w^\downarrow\simeq w/2$ we get
\begin{equation}\label{eq:dipstr}
F_a\left(w^\uparrow+w^\downarrow\right)=\frac{2 w\mathcal{P}_a}{w^\uparrow-w^\downarrow}\simeq 40 k_{\rm B}T\simeq 1.6\times 10^{-19} {\rm J}\;.
\end{equation}
In \cite{manneville01} they achieve concentrations around $n_\Sigma=10^{16}\;{\rm m}^{-2}$. Multiplying this number with the one from Equation (\ref{eq:dipstr}) we find an estimate for the tension
\begin{equation}
\sigma_{\rm dip}\simeq F_a\left(w^\uparrow+w^\downarrow\right)n_\Sigma\simeq 1.6\cdot 10^{-3} {\rm N}/{\rm m}\;.\label{eq:sigmadipest}
\end{equation}

If the membrane is in a floppy state with a large excess of area stored in the fluctuations of its shape, which would for instance be the case if the fluctuations of the membrane were to be measured in a video microscopy experiment, then the membrane can contract its area in response to the increased tension. Thus the true increase in tension will not be as high as indicated by Equation (\ref{eq:sigmadipest}) in that case. Assuming that almost all of the active contribution to the tension is canceled by a contraction we can estimate the relative change in area by $\Delta\alpha=-\sigma_{\rm dip}/K_a$, where $K_a$ is the membrane expansion modulus. A typical value for $K_a$ would be $K_a=0.2\;{\rm N/m}$ \cite{henriksen04}. Using the tension from (\ref{eq:sigmadipest}), we then get a relative decrease in area of around 0.8 percent. To estimate the the small part of the active contribution to the tension which is not canceled we assume that we have a nearly planar membrane sitting on a frame of area $A_0$. Expanding the free energy in Equation (\ref{eq:Fbend}), assuming that the fields contributing the terms indicated by dots have already been integrated out, to second order in deviations $\epsilon=\epsilon(x,y)$ from a planar shape one can find using the equipartition theorem that the fluctuation spectrum of the membrane is
\begin{equation}\label{eq:planespec}
\left<|\epsilon_q|^2\right>=A_0\frac{k_{\rm B}T}{\kappa q^4+\sigma_0 q^2}\;,
\end{equation}
where $\epsilon_q=\int d x\, d y\;e^{i(q_x x+q_y y)}\epsilon(x,y)$ and $q=\sqrt{q_x^2+q_y^2}$. The excess of area stored in these fluctuations can then be calculated as
\begin{align}
\alpha&=\frac{1}{A_0}\int d x\, d y\;\left<(\vec{\nabla}_\|\epsilon)^2/2\right>=\frac{1}{A_0}\int_0^{q_{\rm max}}\frac{q\, d q}{2\pi}\;\frac{q^2}{2}\left<|\epsilon_q|^2\right>\nonumber\\
&=\frac{k_{\rm B}T}{8 \pi \kappa}\log \frac{\kappa q_{\rm max}^2+\sigma_0}{\sigma_0}\simeq -\frac{k_{\rm B}T}{8 \pi \kappa}\log \frac{\sigma_0}{\kappa q_{\rm max}^2}\;,\label{eq:alpsig}
\end{align}
where $\vec{\nabla}_\|=\vec{t}^\alpha\partial_\alpha$ and $q_{\rm max}$ is a cut-off at high wavenumbers where the free energy in Equation (\ref{eq:Fbend}) is not expected to be valid anymore. A cut-off at small wavenumbers is not necessary due to the suppression of the fluctuations by the tension there. Equation (\ref{eq:alpsig}) can be used to obtain $\kappa$ from micropipette experiments by measuring the relative excess area as the tension is changed. In \cite{manneville99} this was used to obtain the value $\kappa\simeq 10 k_{\rm B}T$ for the bending rigidity of a passive lipid-bacteriorhodopsin membrane. When the bacteriorhodopsin molecules were activated by shinning green-yellow light on the membrane then it was found that the change in relative excess area with tension now behaved as
\begin{equation}
\Delta\alpha=-\frac{k_{\rm B}T^{\rm eff}}{8\pi \kappa}\log \frac{\sigma_{{\rm final}}}{\sigma_{{\rm initial}}}\;,\label{eq:alphasigma}
\end{equation}
where the effective temperature introduced here was measured to be $T^{\rm eff}\simeq 2T$. Isolating the tensions we find from Equation (\ref{eq:alphasigma}) that a 0.8 percent decrease in the excess area corresponds to an increase in tension by a factor
\begin{equation}
\frac{\sigma_{{\rm final}}}{\sigma_{{\rm initial}}}\simeq \exp\left(0.008\cdot 8\pi\cdot 10/2\right)\simeq 3\;.
\end{equation}
Thus if the initial tension is much smaller than $\sigma_{\rm dip}$ then the assumption that almost all of the active contribution is canceled by the contraction of the membrane area is not violated by this factor three increase of the initial tension.

Note that the assumption $w^\uparrow-w^\downarrow\simeq w/2$ is a conservative estimate, meaning that if the dipole model in Equation (\ref{eq:symmMBRP}) is appropriate and if $\mathcal{P}_a$ has the value given in Equation (\ref{eq:Paest}), then the estimate in Equation (\ref{eq:dipstr}) is a lower bound on the dipole moment of the individual active proteins. However, if the lower bound is contradicted by experiment, the force-dipole in Equation (\ref{eq:symmMBRP}) could be replaced with a more complex force distribution giving a smaller dipole moment but the same quadrupole moment.

\section{Hydrodynamic interaction between active proteins}\label{sec:diffusion}
As mentioned in the introduction, two experiments on the diffusion dynamics and the interactions of bacteriorhodopsin molecules in their passive and active states were reported in \cite{kahya02}. The experiments showed that the diffusion of the bacteriorhodopsin molecules slows down when they are activated, and that the bacteriorhodopsin molecules tend to cluster more in their active state. Both of these effects can be explained by an attractive force between the bacteriorhodopsin molecules arising in the active state.

A possible mechanism through which this attraction can take place is a coupling between bulk fluid pressure and diffusion currents in the membrane which was also discussed in \cite{lomholt05}. One way to state this mechanism is that there should be a term in the chemical potential of the membrane proteins which is the pressure of the surrounding bulk fluid times the volume of the protein. Thus when the activity of a protein lowers the pressure in its surroundings, Equation (\ref{eq:actpress}), the chemical potential of other molecules in its vicinity will be lowered and therefore there will be an attractive force toward the active protein through this mechanism. However, the lipids in the membrane will also be attracted, and thus a counter-pressure will arise inside the membrane opposing the attraction. But if the proteins stick out of the membrane they will still feel a net attractive potential equal to the lowering of the surrounding pressure times the volume of the part of the protein which is outside the average lipid thickness of the membrane.

To give a rough estimate of the importance of this attractive mechanism, we can compare it with the entropic repulsion of the proteins. Ignoring that proteins can insert themselves with two possible orientations in the membrane we can take the chemical potential of the proteins to be
\begin{equation}
\mu^{\rm p}=k_{\rm B}T\log n_{\rm p}+p\Delta V\;,
\end{equation}
where $\Delta V$ is the volume of the part of the protein which is outside the lipid part of the membrane and $p$ is the pressure of the bulk fluid next to the membrane. Taking the membrane to be planar we get from Equations (\ref{eq:actpress}) and (\ref{eq:symmMBRP}) that $p=p_0-F_a n_p$ where $p_0$ is the pressure in the absence of activity, which we will take to be a constant. Working to first order in deviations from the homogeneous distribution of proteins, we can write the constitutive relation for diffusion as \cite{lomholt05}
\begin{equation}
\frac{\partial n}{\partial t}\simeq D_\alpha\left(\Omega_{\rm p}\partial^\alpha \mu^{\rm p}\right)\simeq D_{\rm p}D_\alpha\partial^\alpha n_{\rm p}\;,
\end{equation}
where $\Omega_{\rm p}$ is a kinetic coefficient for protein diffusion and
\begin{equation}\label{eq:Dp}
D_{\rm p}=\Omega_{\rm p}\left(\frac{k_{\rm B}T}{n_{\rm p}}-F_a\Delta V\right)
\end{equation}
is the corresponding diffusion constant. Thus to estimate the importance of the active contribution relative to the entropic, we can compare the two terms in the parenthesis in Equation (\ref{eq:Dp}). Reusing the protein density $n_{\rm p}\simeq 10^{16}\;{\rm m}^{-2}$ we find
\begin{equation}
\frac{k_{\rm B}T}{n_{\rm p}} \simeq 100\;{\rm nm}^2\cdot k_{\rm B}T\;.
\end{equation}
According to \cite{kimura97} the parts of a bacteriorhodopsin molecule extending into the aqueous surroundings are limited. Let us, however, say that these parts extend approximately 1 nm out. Assuming that the cross-sectional area is approximately $(5\;{\rm nm})^2$ we get a volume of approximately $\Delta V\simeq 25 {\rm nm}^3$. To estimate $F_a$ we can use the estimate in Equation (\ref{eq:dipstr}) and set $w^\uparrow+w^\downarrow\simeq 5\;{\rm nm}$. Doing this we get
\begin{equation}
F_a\Delta V\simeq \frac{40\,k_{\rm B}T}{5\;{\rm nm}}\cdot 25\;{\rm nm}^3 =200\;{\rm nm}^2\cdot k_{\rm B}T\;.\label{eq:numbone}
\end{equation}

The lesson that we can take from the above estimate is that this force-dipole induced hydrodynamic attraction can matter for the clustering and diffusion of bacteriorhodopsin molecules. Working out the importance of this mechanism more precisely is difficult, because it would require knowledge of what additional forces, for instance membrane induced interactions \cite{dan94}, there is between the proteins besides the active and entropic contributions. In the case of bacteriorhodopsin there has to be additional attractive forces, since it is known that bacteriorhodopsin molecules also form clusters in their passive state \cite{kahya02,gulik87}.

It should be mentioned that there is also mechanisms that can actively enhance diffusion. Rotating motors for instance enhance the fluctuations in the local flow around them creating an active contribution to diffusion \cite{lenz04}.

\section{Conclusion}\label{sec:conclusion}
In this paper the force-dipole model of \cite{manneville01} was generalized to arbitrary distributions of forces in the direction transverse to the membrane. The model was then reformulated into a Gibbs formulation, where the activity entered the force-balance equation of the membrane surface directly, instead of being part of the force-balance in the bulk fluids surrounding the membrane. An expression for the active contribution to the force-balance in the Gibbs formulation, $\vec{f}_{\rm act}=D_\alpha\left(T^{\alpha\beta}_{\rm act}\vec{t}_\beta+T_{\rm n,act}^\alpha\vec{n}\right)$, was given through Equations (\ref{eq:tangTact}) and (\ref{eq:normTact}) in the form of a moment expansion of the active force distribution, $F_{\rm act}$. This expression provides a way to study the consequences of the force-dipole model of \cite{manneville01} for non-planar membrane shapes, opening up the possibility of testing the model in a wider class of experiments than theoretical predictions restricted to planar membranes would allow for. 
Some consequences of the activity, however, become immediately apparent in the Gibbs formulation. One is that the dipole moment of the active force distribution $\sigma_{\rm dip}$ gives rise to a modification of the tension of the membrane. The prominent effect of the next order in the moment expansion, the quadrupole moment $Q$, is that it modifies the bending moments of the membrane. The term in the force-balance of the membrane that this results in couples the dynamics of the membrane shape to the protein concentrations, and this term, if the force-dipole model is a proper explanation for the activity induced effect observed in the micropipette experiments \cite{manneville99,girard05}, induces a magnification of primarily the bending rigidity dominated part of fluctuation spectrum \cite{manneville01}, i.e. the $q^{-4}$ behavior of Equation (\ref{eq:planespec}). It should be noted that this modification of the fluctuation spectrum does not correspond simply to a renormalized bending rigidity or a modification of another coupling parameter from the equilibrium free energy. The reason behind this is that the activity does not produce symmetric couplings between the dynamics of the membrane shape and the protein densities. The coupling constants in the free energy produce such symmetric couplings because of Maxwell relations and Onsager's reciprocal relations.

The results obtained in this paper are for a model of the activity that generalizes the force-dipole model of \cite{manneville01}. If one wants to reduce the equations describing the membrane dynamics here to those of \cite{manneville01} one first of all needs to specialize to the force distribution given by Equation (\ref{eq:MBRP}) \footnote{The moment expansion employed in this paper is not a restriction in comparison with \cite{manneville01}, since the same expansion is performed in \cite{manneville01} (without it being mentioned explicitly though).}. Secondly, one has to go to the planar case. Thirdly, thermal noise should be included in the Navier-Stokes equation, Equation (\ref{eq:actNS}) or (\ref{eq:noactNS}) depending on which formulation one begins with. Note that non-thermal noise arising due to the activity is not included in \cite{manneville01}. See \cite{gov04,lomholt05d} for ways to include this. Fourthly, the boundary condition of Equation (\ref{eq:genbound}) should be specified. In \cite{manneville01} effects due to permeation of the membrane are included at the outset, but it is later argued that such effects are negligible for undulations with wavelengths in the regime relevant to experiments. Since bulk fluid motion in the tangential direction of the membrane decouples from the motion of the membrane shape in the planar case (see Equation (\ref{eq:zcompforce}) here) it is enough to choose $\vec{n}\cdot\vec{U}_\pm=\vec{n}\cdot\partial\vec{R}/\partial t$ to obtain the equations of \cite{manneville01} that govern the membrane motion in the experimentally relevant regime. Fifthly, diffusion equations for the proteins in the membrane should be set up. The hydrodynamic attraction proposed in section \ref{sec:diffusion} here can be included simply as a decreased diffusion constant of the active proteins.

Because of the simplified bulk hydrodynamics in the Gibbs formulation, this formulation will constitute a much simpler starting point than the semi-microscopic formulation for further studies of long-wavelength consequences of the force-dipole model. This will be applied in another paper \cite{lomholt05d}, where the fluctuation spectrum of a quasi-spherical vesicle will be derived for the force-dipole model.

\begin{acknowledgments}
I would like to thank Ling Miao, Per Lyngs Hansen and Jean-Fran{\c c}ois Joanny for many illuminating discussions on active membranes. I am grateful to the Villum Kann Rasmussen Foundation for their financial support.
\end{acknowledgments}

\appendix

\section{Solution of the semi-microscopic formulation in the planar case}\label{sec:planact}
In this appendix we will find the solution of the semi-microscopic formulation, Equation (\ref{eq:actNS}), to first order in deviations from a planar shape situated at $z=0$ in a Cartesian coordinate system $(x,y,z)$.

\subsection{Hydrodynamics without active forces}
We need the solution without the active contribution, such that we have a reference when we wish to find the modification due to the activity. To find this solution we will Fourier transform in the plane of the membrane
\begin{equation}
\vec{v}_\pm(x,y,z)=\int\frac{d^2q}{(2\pi)^2} e^{-i(q_x x+q_y y)}\vec{v}_{\pm,\vek{q}}(z)\;,
\end{equation}
and decompose the velocity into its $z$, longitudinal and transverse components as
\begin{equation}
\vec{v}_{\pm,\vek{q}}=v_{{\pm,\vek{q}}}^z\hat{\vec{z}}+v_{{\pm,\vek{q}}}^l\hat{\vec{q}}+v_{{\pm,\vek{q}}}^t\hat{\vec{t}}\;,
\end{equation}
where $\vec{q}=(q_x,q_y,0)=q\vec{\hat q}$ and $\vec{\hat t}$ is a unit vector perpendicular to both of the unit vectors $\vec{\hat z}$ and $\vec{\hat q}$. The solution of (\ref{eq:noactNS}) can then be written
\begin{align}
v_{{\pm,\vek{q}}}^z&=e^{-q|z|}\left[\left(1+q|z|\right)v_{{\pm,\vek{q}}}^z|_{z=0}+iqzv_{{\pm,\vek{q}}}^l|_{z=0}\right]\ ,\label{eq:nosol1}\\
v_{{\pm,\vek{q}}}^l&=e^{-q|z|}\left[\left(1-q|z|\right)v_{{\pm,\vek{q}}}^l|_{z=0}+iqzv_{{\pm,\vek{q}}}^z|_{z=0}\right]\ ,\\
v_{{\pm,\vek{q}}}^t&=e^{-q|z|}v_{{\pm,\vek{q}}}^t|_{z=0}\ ,\\
p^\pm_{{\vek{q}}}&=2\eta qe^{-q|z|}\left[\frac{z}{|z|}v_{{\pm,\vek{q}}}^z|_{z=0}+iv_{{\pm,\vek{q}}}^l|_{z=0}\right]\;.\label{eq:nosol4}
\end{align}

We will also find the forces $\vec{T}^\pm$ of the bulk fluids on the membrane defined in Equation (\ref{eq:Tvecdef}). To this end we need the stress tensors
\begin{equation}
\mathsf{T}^\pm=-p^\pm\mathsf{I}+\eta\left[\vec{\nabla}\vec{v}_\pm+\left(\vec{\nabla}\vec{v}_\pm\right)^T\right]\;.
\end{equation}
The important components at $z=0$ are
\begin{eqnarray}
\mathsf{T}_{zz}^\pm&=&-p_0^\pm\mp{2\eta}\int\frac{d^2q}{(2\pi)^2} e^{-i(q_x x q_y y)} q v^z_{{\pm,\vek{q}}}|_{z=0}\;,\;\;\;\;\\
\mathsf{T}_{zl}^\pm&=&\mp {2\eta}\int\frac{d^2q}{(2\pi)^2} e^{-i(q_x x+q_y y)}q v^l_{{\pm,\vek{q}}}|_{z=0}\ ,\\
T_{zt}^\pm&=&\mp {\eta}\int\frac{d^2q}{(2\pi)^2} e^{-i(q_x x+q_y y)}q v^t_{{\pm,\vek{q}}}|_{z=0}\;,
\end{eqnarray}
and they give the forces on the membrane
\begin{align}
\vec{T}^\pm_{\pr,{\vek{q}}}\cdot\vec{\hat q}=&- {2\eta}q v^l_{{\pm,\vek{q}}}|_{z=0}\;,\\
\vec{T}^\pm_{\pr,{\vek{q}}}\cdot\vec{\hat t}=&- {\eta}q v^t_{{\pm,\vek{q}}}|_{z=0}\;,\\
\left(\vec{T}^\pm\cdot\vec{n}\right)_{\vek{q}}=&\mp p_0^\pm\left(2\pi\right)^2\delta\left(q_x\right)\delta\left(q_y\right)-2\eta q v^z_{{\pm,\vek{q}}}|_{z=0}\;,\label{eq:zcompforce}
\end{align}
where $\vec{T}^\pm_{\pr}\equiv \vec{T}^\pm\cdot\vec{t}_\alpha \vec{t}^\alpha$ is the projection of the forces onto the tangent space of the membrane.

\subsection{Hydrodynamics with active forces}
We will consider the contribution to Equation (\ref{eq:actNS}) from a single ``layer'' of active force
\begin{equation}
\vec{F}_{{\rm act},h}=d h \int_{\rm M} d A \;F_{\rm act}\vec{n}\delta^3\left(\vec{r}-\left(\vec{R}+h\vec{n}\right)\right)\;.
\end{equation}
Expanding to first order in deviations $\epsilon$ from the planar shape we get
\begin{align}
\vec{F}_{{\rm act},h}=&F_{\rm act} d h\left\{\vec{\hat z}\left(1+h\laplace\epsilon-\epsilon\partial_z\right)-\vec{\nabla}_\pr \epsilon\right\}\delta\left(z- h\right)\;,
\end{align}
where $\vec{\nabla}_\pr=\vec{t}^\alpha\partial_\alpha$. Using this force in the Navier-Stokes Equation (\ref{eq:actNS}), together with the boundary condition that the velocity should vanish at $z=0$, we get for $|z|<|h|$
\begin{align}
p^\pm_{\vek{q}}=&{\tilde F}_{{\rm act},h,\vek{q}}e^{-q |h|}\big[\mp \cosh(qz)-q h e^{-q |z|}\big]\;,\label{eq:sol1}\\
v_{\pm,\vek{q}}^z=&\frac{1}{2\eta} {\tilde F}_{{\rm act},h,\vek{q}}e^{-q |h|}\big[-z e^{-q |z|}q h \nonumber\\
&\;\;+(q^{-1}+|h|)\sinh(q |z|)-|z|\cosh(q |z|)\big]\;,\\
v_{\pm,\vek{q}}^l=&\mp\frac{i}{2\eta}{\tilde F}_{{\rm act},h,\vek{q}}e^{-q |h|}\big[(h-z)\sinh(q z)+q h z e^{-q |z|}\big]
\end{align}
and for $|z|>|h|$
\begin{align}
p^\pm_{\vek{q}}=& {\tilde F}_{{\rm act},h,\vek{q}}e^{-q |z| }\big[\left(\sinh(q h)-q h e^{-q |h|}\right)\big]\;,\\
v_{\pm,\vek{q}}^z=&\frac{1}{2\eta} {\tilde F}_{{\rm act},h,\vek{q}}e^{-q |z| }\big[-z (q h e^{-q |h|}-\sinh(q h))\nonumber\\
&\;\;\pm (q^{-1}\sinh(q h)-h\cosh(q h))\big]\;,\\
v_{\pm,\vek{q}}^l=&\mp\frac{i}{2\eta}{\tilde F}_{{\rm act},h,\vek{q}}e^{-q |z| }\big[(h-z)\sinh(q h)+q h e^{-q |h|}z \big]\label{eq:sol3}
\end{align}
where
\begin{align}
{\tilde F}_{{\rm act},h}&=F_{\rm act} d h\left(1+h \laplace\epsilon\right)\;,
\end{align}
and it is assumed that $h$ and $z$ are positive for the ``$+$''-part and negative for the ``$-$''-part. For $v_{\pm,\vek{q}}^t$ we simply have $v_{\pm,\vek{q}}^t=0$ everywhere.

We can write Equations (\ref{eq:sol1}) to (\ref{eq:sol3}) as
\begin{align}
v_{{\pm,\vek{q}}}^z=&e^{-q|z|}\left[\left(1+q|z|\right){\bar v}_{{\pm,\vek{q}}}^z|_{z=0}+iqz{\bar v}_{{\pm,\vek{q}}}^l|_{z=0}\right]\label{eq:exsol1}\nonumber\\
&+v_{{\pm,\vek{q}}}^{z,{\rm extra}}\;,\\
v_{{\pm,\vek{q}}}^l=&e^{-q|z|}\left[\left(1-q|z|\right){\bar v}_{{\pm,\vek{q}}}^l|_{z=0}+iqz{\bar v}_{{\pm,\vek{q}}}^z|_{z=0}\right]\nonumber\\
&+v_{{\pm,\vek{q}}}^{l,{\rm extra}}\;,\\
p^\pm_{\vek{q}}=&2\eta qe^{-q|z|}\left[\frac{z}{|z|}v_{{\pm,\vek{q}}}^z|_{z=0}+iv_{{\pm,\vek{q}}}^l|_{z=0}\right]+p_{{\vek{q}}}^{\pm,{\rm extra}}\;,\label{eq:exsol3}
\end{align}
where
\begin{align}
{\bar v}_{{\pm,\vek{q}}}^z|_{z=0}&=\pm\frac{1}{2\eta}{\tilde F}_{{\rm act},h,\vek{q}}\left(q^{-1}\sinh(qh)-h\cosh(q h)\right)\;,\label{eq:bound1}\\
{\bar v}_{{\pm,\vek{q}}}^l|_{z=0}&=\mp i\frac{1}{2\eta}{\tilde F}_{{\rm act},h,\vek{q}}h\sinh(q h)\;,\label{eq:bound2}
\end{align}
and the extra part is zero for $|z|>|h|$ and
\begin{align}
v_{{\pm,\vek{q}}}^{z,{\rm extra}}=&\pm\frac{{\tilde F}_{{\rm act},h,\vek{q}}}{2\eta}\big[\left(h-z\right)\cosh(q(z-h))\nonumber\\
&\;\;+q^{-1}\sinh(q(z-h))\big]\;,\\
v_{{\pm,\vek{q}}}^{l,{\rm extra}}=&\mp i\frac{{\tilde F}_{{\rm act},h,\vek{q}}}{2\eta}\left[\left(h-z\right)\sinh(q(z-h))\right]\;,\\
p_{{\vek{q}}}^{\pm,{\rm extra}}=&\mp{{\tilde F}_{{\rm act},h,\vek{q}}}\cosh(q(z-h))\;,
\end{align}
when $|z|<|h|$. This extra part gives a force on the membrane which is
\begin{align}
\vec{T}_{{\vek{q}}}^{\pm,{\rm extra}}\cdot\vec{\hat z}&={{\tilde F}_{{\rm act},h,\vek{q}}}\left(\cosh(q h)-q h \sinh(q h)\right)\;,\label{eq:exforce1}\\
\vec{T}_{\pr,{\vek{q}}}^{\pm,{\rm extra}}\cdot\vec{\hat q}&=-{i}{{\tilde F}_{{\rm act},h,\vek{q}}}q h\cosh(q h)\;.\label{eq:exforce2}
\end{align}

Equations (\ref{eq:exsol1}) to (\ref{eq:exsol3}) should be compared with Equations (\ref{eq:nosol1}) to (\ref{eq:nosol4}). Doing this we see that we get the same behavior in the bulk for $|z|>|h|$ if remove the active force from the Navier-Stokes equation and instead use the boundary conditions in Equations (\ref{eq:bound1}) and (\ref{eq:bound2}). To second order in $h$ these boundary conditions are
\begin{align}
{\bar v}_{{\pm,\vek{q}}}^z|_{z=0}&=0\;,\\
{\bar v}_{{\pm,\vek{q}}}^l|_{z=0}&=\mp iq\frac{1}{2\eta}F_{{\rm act},h,{\vek{q}}}d h\,h^2\;,
\end{align}
which agrees with the boundary condition in the Gibbs formulation, Equation (\ref{eq:actbc}).

If the Gibbs formulation is correct, then Equations (\ref{eq:exforce1}) and (\ref{eq:exforce2}) should match the active force $\vec{f}_{\rm act}$ from section \ref{sec:excess}. Expanding Equations (\ref{eq:exforce1}) and (\ref{eq:exforce2}) to second order in $h$ we find
\begin{align}
\left(\vec{T}^{+,{\rm extra}}_{\vek{q}}+\vec{T}^{-,{\rm extra}}_{\vek{q}}\right)\cdot\vec{\hat z}=&F_{{\rm act},h,\vek{q}}d h+2 F_{{\rm act},h,\vek{q}=0} d h\,h H_{\vek{q}}\nonumber\\
&-\frac{1}{2}q^2F_{{\rm act},h,{\vek{q}}}d h\,h^2\;,\label{eq:gibbs1}\\
\left(\vec{T}^{+,{\rm extra}}_{\pr,\vek{q}}+\vec{T}^{-,{\rm extra}}_{\pr,\vek{q}}\right)\cdot\vec{\hat q}=&-iq F_{{\rm act},h,{\vek{q}}}d h\,h\nonumber\\
&-2iq H_{\vek{q}}F_{{\rm act},h,\vek{q}=0} d h\,h^2 \label{eq:gibbs2}\;,
\end{align}
and again agreement with the Gibbs formulation, Equations (\ref{eq:tangTact}) and (\ref{eq:normTact}), is found when it is recalled that the monopole moment of the force $F_{\rm act}$ vanishes.

\bibliography{../mycites/mycites}

\end{document}